\documentclass[journal]{IEEEtran}
\usepackage{amsmath,amsfonts}
\usepackage{algorithmic}
\usepackage{algorithm}
\usepackage{array}
\usepackage[caption=false,font=footnotesize,labelfont=rm,textfont=rm]{subfig}
\usepackage{textcomp}
\usepackage{stfloats}
\usepackage{url}
\usepackage{verbatim}
\usepackage{graphicx}
\usepackage{cite}
\usepackage{multirow}
\usepackage{booktabs}
\usepackage{makecell}
\usepackage{tablefootnote}
\usepackage{diagbox}
\usepackage{hhline}
\usepackage{pifont}
\usepackage{threeparttable}
\usepackage{xstring}
\usepackage{tabularray}
\usepackage[table]{xcolor}

\definecolor{rowgray}{gray}{0.93}
\ifCLASSINFOpdf
\else
   \usepackage[dvips]{graphicx}
\fi
\usepackage{url}

\usepackage{graphicx}

\makeatletter
\newcommand{\ssymbol}[1]{^{\@fnsymbol{#1}}}

\newif\ifrevision
\revisionfalse

\newcommand{\tabvspace}{\vspace{-5pt}}

\ifrevision
\def\cblue{\textcolor{blue}}

\else
\def\cblue#1{#1}

\fi

\begin{document}

\title{Echo-Aware Modulation for Compact-Latent Frequency-Time Modeling in Lightweight Acoustic Echo Cancellation}

\author{Ye Ni, Ruiyu Liang, Qingyun Wang, Kai Xie, Cairong Zou, Bj\"orn W.\ Schuller, \IEEEmembership{Fellow, IEEE} 
\thanks{This work was supported in part by the National Natural Science Foundation of China under grant number 61871213, and in part by the China Scholarship Council under Grant 202406090186. (Corresponding authors: Ruiyu Liang.)}
\thanks{Ye Ni, Ruiyu Liang and Cairong Zou are with the School of Information Science and Engineering, Southeast University, Nanjing 210096, China (email: niye@seu.edu.cn; liangry@njit.edu.cn; cairong@seu.edu.cn).}
\thanks{Qingyun Wang is with the School of Communication Engineering, Nanjing Institute of Technology, Nanjing 211167, China (e-mail: wangqingyun@njit.edu.cn).}
\thanks{Kai Xie is with Center of Intelligent Acoustics and Immersive Communications, Northwestern Polytechnical University, China (e-mail: xiekai@mail.nwpu.edu.cn).}
\thanks{Björn W. Schuller is with CHI -- the Chair of Health Informatics, Technical University of Munich, 81675 Munich, Germany, and also with GLAM -- the Group on Language, Audio, \& Music, Imperial College London, SW7 2AZ London, U.\,K. (e-mail: schuller@tum.de).}
}

\markboth{Journal of \LaTeX\ Class Files, Vol. 14, No. 8, August 2015}
{Shell \MakeLowercase{\textit{et al.}}: Bare Demo of IEEEtran.cls for IEEE Journals}
\maketitle

\begin{abstract}
	
\cblue{
Existing lightweight acoustic echo cancellation (AEC) systems often combine linear AEC with Bark-domain DNN-based suppression to lower the computational footprint.
In such systems, downsampling layers further compress the input features into a compact bottleneck representation, but this compression weakens frequency-time modeling capacity and degrades performance.
To mitigate this limitation, we propose MSA-EchoLite, a lightweight Bark-domain AEC framework with an asymmetric dual-branch encoder and an echo-aware frequency-time modulation (EAM) module.
The EAM module enriches the compressed bottleneck representation by modeling discrepancy and correlation cues between the dual-branch microphone and echo-related latent features.
Experimental results show that the Bark-domain variant of MSA-EchoLite offers a better performance-complexity trade-off than its frequency-domain counterpart but is more sensitive to feature compression.
With only 26.1\% additional FLOPs over its non-EAM Bark-domain variant, its EAM-enhanced version achieves 99.1\% of the PESQ of the frequency-domain counterpart, which requires nearly twice the FLOPs, and even surpasses it in SDR.
Overall, MSA-EchoLite outperforms state-of-the-art lightweight AEC models while using only 0.2\,M parameters and 100\,M FLOPs/s.
}
\end{abstract}


\IEEEpeerreviewmaketitle

\section{Introduction}

\IEEEPARstart{A}{coustic} echo cancellation (AEC) is essential for hands-free communication, where loudspeaker playback is captured by the microphone and causes undesired echo. 
Since mobile devices typically have limited computational resources, AEC systems must suppress echo under strict latency, memory, and complexity constraints, making lightweight yet effective models critical for deployment.

Recent DNN-based AEC methods can be broadly categorized into fully neural models~\cite{sun2023multi,ristea2023deepvqe,chen2023progressive,khanagha2024interference,ni2024msa,11460730} and hybrid systems~\cite{zhang2022multi,zhang2022deep,yang2023low,shetu2024hybrid,10715013,li2025echofree,10896787} that combine classical adaptive filtering with DNN-based residual echo suppression. 
Despite different system designs, mainstream neural AEC models~\cite{zhang2022multi,ristea2023deepvqe,khanagha2024interference,ni2024msa} commonly employ convolutional encoder-decoder architectures with frequency-time (FT) modeling bottlenecks.
Such designs can effectively exploit spectral and temporal dependencies, but usually rely on high-dimensional input features, full-band mask estimation, and relatively expensive bottlenecks.
This makes them less suitable for resource-constrained devices.

To reduce the footprint, recent studies have explored compact perceptual-domain representations, such as Bark and equivalent rectangular bandwidth (ERB) scales, which reduce both feature and mask dimensions. 
The Bark-scale postfilter in~\cite{seidel2024efficient} uses fully connected and stacked gated recurrent unit (GRU) layers for residual echo and noise suppression, while EchoFree~\cite{li2025echofree} combines U-Net-based Bark-gain estimation with semantic loss. 
CAGCRN~\cite{wang2025cagcrn} further adopts ERB-band compression with cross-attention alignment and gated convolutional recurrent network (CRN) modeling for joint echo and noise suppression.
Although these methods reduce computational cost, the performance gap between compact-domain and frequency-domain modeling remains underexplored. 
Moreover, compact-domain compression introduces a new challenge that has received limited attention in existing lightweight AEC systems: conventional frequency-time modeling~\cite{gulati2020conformer,luo2020dual,le2021dpcrn} becomes less effective in compact latent spaces, where spectral details and microphone-reference interaction cues are weakened, degrading echo-speech discrimination.

\cblue{Following the hybrid AEC paradigm, in this paper, we propose MSA-EchoLite, a lightweight U-Net-based framework that combines a conventional adaptive filter with a neural suppressor to reduce the learning burden of echo suppression.
The neural suppressor operates on compressed Bark-domain features, adopts an asymmetric dual-branch encoder to further downsample microphone and echo-related representations into a compact bottleneck representation, and predicts a Bark-domain suppression gain to reduce computational complexity.
To compensate for the limited feature capacity caused by Bark-domain compression and successive encoder downsampling, we propose an echo-aware modulated frequency-time module that captures discrepancy and relation cues between the two latent representations within the bottleneck.
We also systematically compare Bark- and frequency-domain modeling within the same framework, revealing their performance-complexity trade-offs and showing that Bark-domain modeling improves efficiency but is more sensitive when the bottleneck feature dimension is reduced.
}

\begin{figure*}[t]
	\centering
	\includegraphics[width=0.85\linewidth]{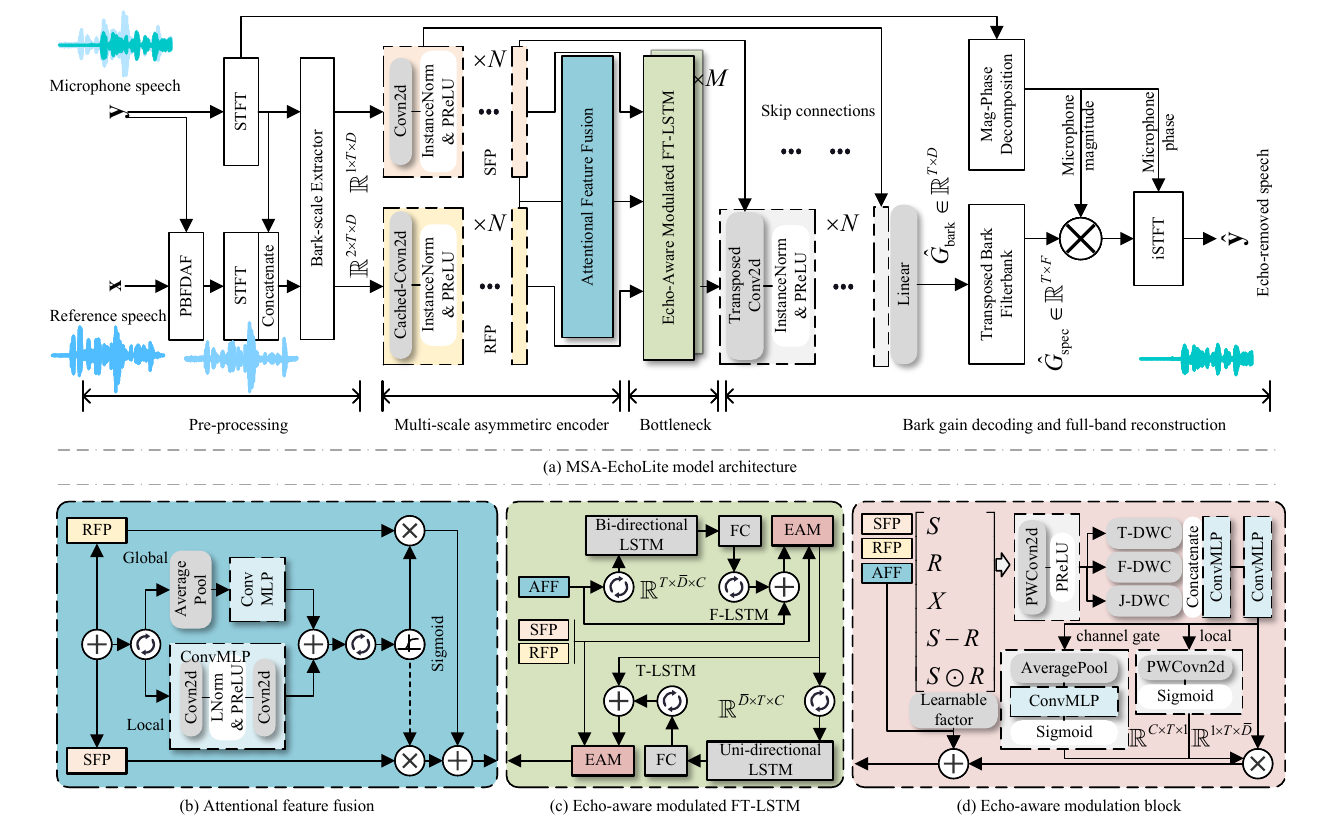}
\caption{(a) Overall architecture of the proposed MSA-EchoLite. (b)--(d) Detailed structures of its constituent modules.}
	\label{fig:model_arch}
\end{figure*}

\section{METHODOLOGY}

\subsection{Problem formulation}

In an AEC system, the far-end reference signal $x(n)$ is rendered by a loudspeaker and captured by the near-end microphone:
\cblue{\begin{equation}
	\begin{aligned}
		y(n)&=d(n)+s(n)+v(n), \\
		d(n) &= h(n) * g\bigl(x(n-\Delta)\bigr),
	\end{aligned}
\end{equation}
where $d(n)$ denotes the echo component, $s(n)$ and $v(n)$ denote the near-end speech and background noise, respectively, $h(n)$ is the acoustic echo path, $g(\cdot)$ represents the nonlinear mapping induced by the loudspeaker and playback chain, and $\Delta$ denotes the system delay.}
The goal of AEC is to estimate $s(n)$ from $y(n)$ by suppressing the echo components.
In this work, we use a partitioned block frequency-domain adaptive filter (PBFDAF)~\cite{pbfdaf} to estimate the linear echo $\hat{d}_{\mathrm{L}}(n)$, which, together with $y(n)$, is fed into MSA-EchoLite $\mathcal{M}(y(n),\hat{d}_{\mathrm{L}}(n);\theta)$ for estimating the speech extraction mask.

\tabvspace
\subsection{Asymmetric Microphone-Echo Encoder}

As shown in Fig.\,\ref{fig:model_arch}(a), MSA-EchoLite builds on the asymmetric dual-path encoding strategy of our previous work~\cite{ni2024msa} and redesigns this strategy as a lightweight Bark-domain encoder using a stack of downsampling convolutional blocks.
The speech feature path (SFP) preserves near-end speech details from the microphone signal, while the reference feature path (RFP) models microphone-reference correlations for echo suppression.
\cblue{Specifically, each encoder stage uses a $1\times5$ convolution in the SFP and a $3\times5$  convolution with causal temporal padding in the RFP, followed by instance normalization and PReLU activation.
The larger temporal kernel in the RFP helps capture short-term microphone-reference correlations.}
These two representations are then fused by an attentional feature fusion block before the bottleneck. 


\tabvspace
\subsection{Echo-Aware Modulated Frequency-Time LSTM Block}

\cblue{
From the data processing inequality~\cite[Section~2.8]{cover2006elements}, since the compressed representation $Z=f(X)$ is obtained only from the original feature $X$, $Z$ cannot contain more task-relevant information about the target $Y$ than $X$, i.e., $I(Z;Y)\leq I(X;Y)$. 
Therefore, aggressive Bark-domain compression and encoder downsampling may discard frequency-time details and microphone-echo interaction cues, leading to performance degradation.}

To mitigate this, we embed EAM after both the frequency and temporal modeling stages of the FT-LSTM bottleneck, as shown in Fig.~\ref{fig:model_arch}(c), with the EAM structure detailed in Fig.~\ref{fig:model_arch}(d).
Let $X, S, R \in \mathbb{R}^{C\times T\times \bar{D}}$ denote the fused, microphone, and echo-related latent representations, respectively.
EAM first forms an echo-aware interaction representation and projects it with a point-wise convolution (PWC) $\mathcal{P}_{p}(\cdot)$:
\begin{equation}
	U=\delta\left(\mathcal{P}_{p}(\mathrm{Concat}(S,R,X,S-R,S\odot R))\right),
\end{equation}
where $\odot$ denotes element-wise multiplication and $\delta(\cdot)$ denotes the PReLU activation.
The difference term captures point-wise discrepancies between the microphone and reference paths, while the product term models their local correlation.
\begin{table}[t]
\caption{Compact-latent comparison of Bark-domain baselines, EAM-enhanced models, and frequency-domain variants. }
\label{tab:compact_ablation}
\centering
\renewcommand{\arraystretch}{1.0}
\setlength{\tabcolsep}{2.8pt}
\resizebox{0.98\columnwidth}{!}{%
\begin{threeparttable}
\begin{tabular}
{l*{3}{>{\centering\arraybackslash}p{0.8cm}}*{5}{>{\centering\arraybackslash}p{0.6cm}}}
\toprule
\multirow{2}{*}{Config} & \multirow{2}{*}{Model} & \#Param & FLOPs & PESQ & ESTOI & SDR & Echo & Deg \\
&       & (M)     & (M/s) &    WB    & (\%)  & (dB) &     MOS    &     MOS   \\
\midrule
\rowcolor{rowgray}
C32H32  & Bark     & 0.10 & 54.68  & 1.97 & 78.95 & 11.13 & 4.47 & 3.81 \\
\rowcolor{rowgray}
& +EAM$\ssymbol{2}$     & 0.15 & 75.71  & 2.01 & 79.35 & 11.46 & 4.51 & 3.86 \\
\rowcolor{rowgray}
& Freq     & 0.16 & 132.65 & 2.03 & 79.77 & 11.29 & 4.57 & 3.88 \\
\midrule
C32H64  & Bark     & 0.15 & 80.83  & 2.02 & 79.67 & 11.59 & 4.50 & 3.84 \\
& +EAM     & 0.20 & 101.90 & 2.10 & 80.68 & 11.93 & 4.54 & 3.94 \\
& Freq     & 0.22 & 196.26 & 2.12 & 80.93 & 11.71 & 4.60 & 3.96 \\
\midrule
\rowcolor{rowgray}
C64H64  & Bark     & 0.25 & 126.93 & 2.15 & 81.14 & 12.28 & 4.57 & 3.99 \\
\rowcolor{rowgray}
& +EAM     & 0.44 & 209.07 & 2.19 & 81.55 & 12.47 & 4.58 & 4.00 \\
\rowcolor{rowgray}
& Freq     & 0.31 & 307.12 & 2.19 & 81.67 & 12.05 & 4.63 & 4.03 \\
\midrule
C32H128 & Bark     & 0.34 & 165.72 & 2.17 & 81.44 & 12.24 & 4.57 & 3.98 \\
& +EAM     & 0.39 & 186.80 & 2.19 & 81.69 & 12.39 & 4.59 & 4.00 \\
& Freq     & 0.41 & 402.44 & 2.23 & 82.27 & 12.34 & 4.64 & 4.04 \\
\midrule
\rowcolor{rowgray}
C64H128 & Bark     & 0.48 & 229.89 & 2.25 & 82.33 & 12.70 & 4.60 & 4.05 \\
\rowcolor{rowgray}
& +EAM     & 0.67 & 312.03 & 2.28 & 82.74 & 12.80 & 4.61 & 4.06 \\
\rowcolor{rowgray}
& Freq     & 0.55 & 557.17 & 2.30 & 83.14 & 12.68 & 4.65 & 4.08 \\
\bottomrule
\end{tabular}%
\begin{tablenotes}
\item[] $\ssymbol{2}$ +EAM denotes the Bark-domain model enhanced by EAM.
\end{tablenotes}
\end{threeparttable}
}
\vspace{-0.2cm}
\end{table}

Three depthwise convolution (DWC) branches are applied to $U$ to obtain $U^{\mathrm{J}}$, $U^{\mathrm{T}}$, and $U^{\mathrm{F}}$ with receptive fields $3\times3$, $3\times1$, and $1\times3$, capturing joint time-frequency, temporal-only, and frequency-only interactions, respectively.
The three branch outputs are then aggregated by:
\begin{equation}
	Z=
	\mathcal{P}_{o}
	\left(
	\mathrm{Concat}(U^{\mathrm{J}},U^{\mathrm{T}},U^{\mathrm{F}})
	\right),
\end{equation}
where $\mathcal{P}_{o}(\cdot)$ denotes the output projection composed of two ConvMLP modules.
EAM controls the interaction using a channel gate $g_{\mathrm{ch}}\in\mathbb{R}^{C\times T\times 1}$ and a local gate $g_{\mathrm{loc}}\in\mathbb{R}^{1\times T\times \bar{D}}$, where $\bar{D}$ is the downsampled latent dimension. 
The channel gate is obtained by averaging over the latent Bark dimension, followed by ConvMLP and sigmoid, while the local gate is generated by PWC and sigmoid. 
The output is computed as:
\begin{equation}
	X_{\mathrm{out}} = X + \alpha\cdot g_{\mathrm{ch}}\odot g_{\mathrm{loc}}\odot Z,
\end{equation}
where $\alpha$ is a learnable scaling factor.

\tabvspace
\subsection{Bark-Domain Training Objective}

Given a spectrum $Z(t,k)$ and Bark filterbank weights $B(k,b)$ for Bark band $b$, the Bark-domain power is computed as:
\begin{equation}
	P_{\mathrm{B}}^{Z}(t,b)
	= \sum_{k=0}^{F-1} |Z(t,k)|^2 B(k,b),
	\quad b=0,\dots,B-1.
\end{equation}
The oracle gain $G_{\mathrm{bark}}^{\star}(t,b)
=
\sqrt{
	\frac{P_{\mathrm{B}}^{S}(t,b)}
	{P_{\mathrm{B}}^{Y}(t,b)+\epsilon}
}$ can be mapped back to the full-band domain through the transposed Bark filterbank and used for reconstruction:
\begin{equation}
	\hat{S}(t,k)
	=
	\left(
	\sum_{b=0}^{B-1} G_{\mathrm{bark}}^{\star}(t,b) B(k,b)
	\right)
	|Y(t,k)|e^{j\angle Y(t,k)}.
\end{equation}

\tabvspace
\subsection{Loss Function}

The predicted Bark gain is not directly supervised; instead, the model is optimized using end-to-end reconstruction losses.
The total loss combines multi-resolution STFT~\cite{defossez2020real}, PMSQE~\cite{martin2018deep}, and the magnitude, time-domain, complex spectral, and anti-wrapping phase losses following~\cite{lu2023mp}:
\begin{equation}
	\begin{aligned}
		\mathcal{L}_{\text{total}}=& \mathcal{L}_{\text{STFT}}
		+ 0.3\mathcal{L}_{\text{PMSQE}}
		+ 0.45\mathcal{L}_{\text{Mag}} \\
		&+ 0.2\mathcal{L}_{\text{Time}}
		+ 0.05\mathcal{L}_{\text{Com}}
		+ 0.3\mathcal{L}_{\text{Pha}} .
	\end{aligned}
\end{equation} 


\section{EXPERIMENTS AND ANALYSIS}

\begin{figure}[t]
	\setlength{\abovecaptionskip}{-0.2cm}
	\centering
	\includegraphics[width=0.98\linewidth]{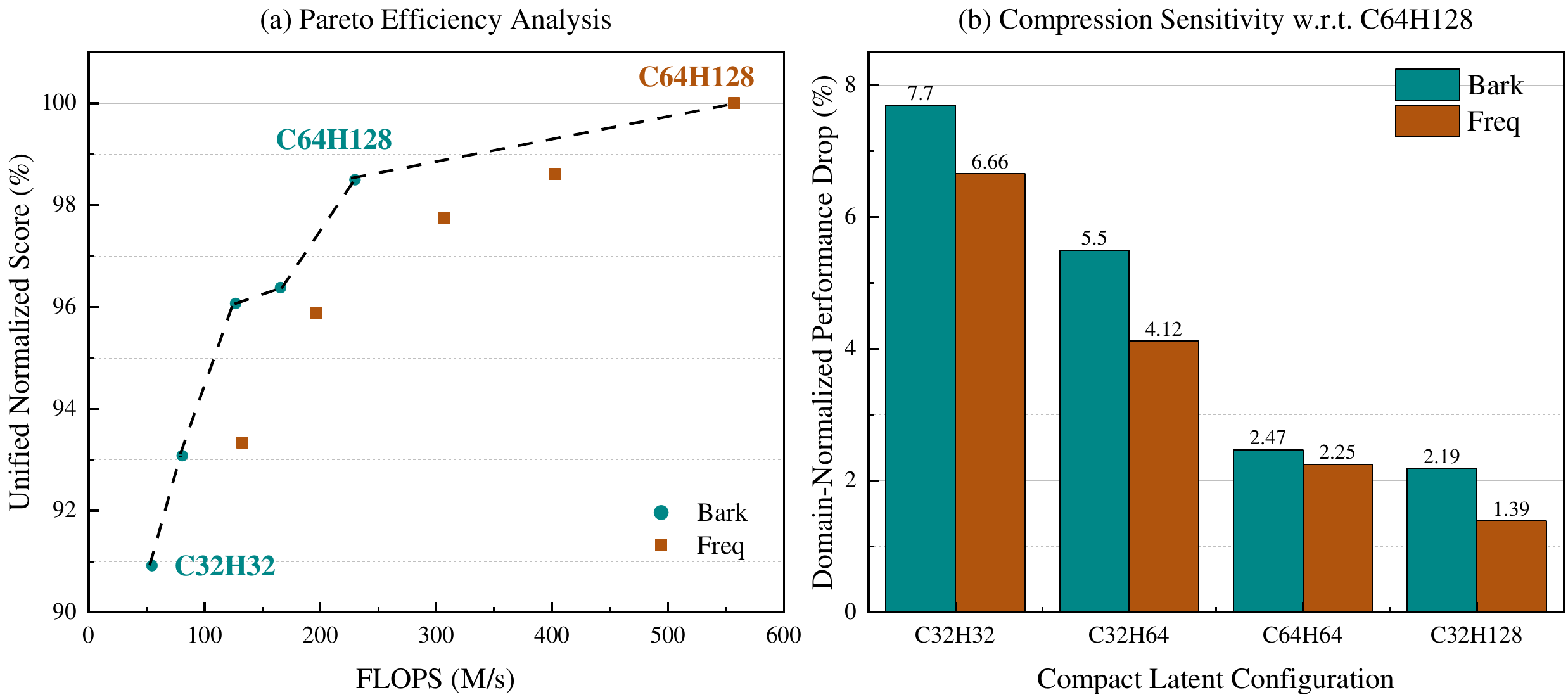}
	\caption{Comparison of Bark- and frequency-domain modeling under compact latent configurations. 
		(a) Pareto efficiency analysis based on a unified normalized score with Freq-C64H128 as reference. 
		(b) Domain-normalized performance drop using the corresponding C64H128 model as reference. 
	}
	\label{fig:compact_latent}
	\vspace{-0.2cm}
\end{figure}

\begin{table}[t]
\cblue{
\caption{Performance of EAM with different numbers of Bark bands.}
\label{tab:compact_ablation_2}
\centering
\renewcommand{\arraystretch}{1.0}
\setlength{\tabcolsep}{2.8pt}
\resizebox{0.98\columnwidth}{!}{%
\begin{threeparttable}
\begin{tabular}
{l*{2}{>{\centering\arraybackslash}p{0.8cm}}*{5}{>{\centering\arraybackslash}p{0.6cm}}}
\toprule
\multirow{2}{*}{Bark bands} & \#Param & FLOPs & PESQ & ESTOI & SDR & Echo & Deg \\
&     (M)     & (M/s) &    WB    & (\%)  & (dB) &     MOS    &     MOS   \\
\midrule
\rowcolor{rowgray}
B48    & 0.14&45.35&1.90&77.71&11.06&4.31&3.77\\
\rowcolor{rowgray}
\  +EAM &   0.19&57.45&1.92&78.05&11.19&4.34&3.81\\
\midrule
B64    &0.14&56.91&1.97&78.79&11.44&4.40&3.85\\
\  +EAM & 0.19&72.00&2.00&79.16&11.58&4.41&3.88\\
\midrule
\rowcolor{rowgray}
B100    &0.15&80.83  & 2.02 & 79.67 & 11.59 & 4.50 & 3.84 \\
\rowcolor{rowgray}
\  +EAM    & 0.20 & 101.90 & 2.10 & 80.68 & 11.93 & 4.54 & 3.94 \\
\bottomrule
\end{tabular}%
\end{threeparttable}
}
\vspace{-0.2cm}
}
\end{table}

\subsection{Dataset}

The training data are constructed from the AEC Challenge~\cite{icassp2023} and DNS Challenge wide-band datasets~\cite{reddy2020interspeech}. 
We synthesize 100 hours of double-talk (DT) and 30 hours of near-end single-talk (NE) data, with each sample lasting 10 seconds. 
For the NE scenario, clean and noisy speech are generated with probabilities of 0.7 and 0.3, respectively, and the SNR is uniformly sampled from 10 to 30\,dB. 
Samples with an activation ratio below 0.6 or average energy below $-30$\,dB are discarded. 
For the DT scenario, reference-echo pairs from the AEC Challenge are mixed with NE speech using a signal-to-echo ratio uniformly sampled from $-15$ to 18\,dB. 
We use the simulated test set for ablation studies, where clean references are available for objective evaluation. 
To examine generalization in realistic conditions, we further compare MSA-EchoLite with representative baselines on the real-recorded AEC Challenge blind test sets.

\tabvspace
\subsection{Experiment Settings}

\subsubsection{Implementation Details}
All models are trained under the same protocol for fair comparison. 
The frame/hop sizes are 512/256.
Models are trained for 60 epochs using Adam with an initial learning rate of $5\times10^{-4}$, which is halved every 15 epochs.
For compact-latent FT analysis, we vary Bark bands $B\in\{48,64,100\}$, bottleneck channels $C\in\{32,64\}$, and hidden size $H\in\{32,64,128\}$.

\subsubsection{Baselines}

\begin{table}[t]
\caption{Evaluation on the real blind test sets of the AEC Challenge using ITU-T P.831 subjective ratings and ERLE.}
\label{table_ablation_study_dns}
\centering
\resizebox{\columnwidth}{!}{%
\begin{threeparttable}
\begin{tabular}{
l
>{\centering\arraybackslash}p{0.4cm}
>{\centering\arraybackslash}p{0.6cm}
>{\centering\arraybackslash}p{0.3cm}
*{8}{>{\centering\arraybackslash}p{0.28cm}}
>{\centering\arraybackslash}p{0.65cm}
}
\toprule
\multirow{2.5}{*}{Models} & 
\#Pa.&
FLOPs&
\cblue{RTF}&
\multicolumn{4}{c}{EchoMOS} & 
\multicolumn{4}{c}{DegMOS}  &
\multirow{2.5}{*}{ERLE}   \\
\cmidrule(lr){5-8}
\cmidrule(lr){9-12}
& (M) & (M/s) &\cblue{(\%)}
& Cln & Nsy & DT & FE 
& Cln & Nsy & DT & NE  & \\
\midrule
PBFDAF~\cite{pbfdaf}&-&-&-&2.61&2.65&2.90&2.36&\bf 3.91&\bf3.80&\bf3.87&3.82& 2.84\\
\midrule
DTLN-AEC~\cite{westhausen2021acoustic} &1.33&83.90&\cblue{0.47}&4.01&3.78&3.94&3.85&3.34&3.05&3.05&3.48 &23.42\\
DeepVQE-S~\cite{ristea2023deepvqe}&0.64&287.31&\cblue{5.21}&4.22&3.91&4.10&4.03&3.79&3.58&3.61&3.82&22.78\\
EchoFree~\cite{li2025echofree} &0.30&27.71&\cblue{0.70}&4.32&4.02&4.26&4.08&3.91&3.74&3.76&\bf3.95&25.37\\
MSA-DPCRN~\cite{ni2024msa}&0.17&297.38&\cblue{2.56}&4.25&3.95&4.10&4.10&3.67&3.37&3.38&3.80&\bf 27.86\\
\midrule
MSA-EchoLite (ours)  & 0.20&101.90&\cblue{1.41}& \bf4.39 & \bf4.15 &\bf 4.28 & \bf4.26 & 3.85 &3.56 & 3.63 &3.85&27.69 \\
\bottomrule
\end{tabular}%
\begin{tablenotes}
\item[1] All DNN baselines are retrained using the same training data and loss function as ours.
\item[2] Cln, Nsy, DT, FE, and NE denote clean, noisy, double-talk, far-end single-talk, and near-end single-talk scenarios, respectively.
\item[3] RTF is reported as a percentage, measured on a single Intel Xeon Gold 6426Y CPU thread.
\end{tablenotes}
\end{threeparttable}
}
\vspace{-10pt}
\end{table}

We consider several representative baselines: 1) DTLN-AEC~\cite{westhausen2021acoustic}, a lightweight two-stage recurrent model operating in spectral and time domains; 2) DeepVQE~\cite{ristea2023deepvqe}, a high-capacity end-to-end framework with substantially higher computational complexity, serving as an upper-bound comparison; 3) MSA-DPCRN~\cite{ni2024msa}, a multi-scale dual-path CRN with separate microphone and reference encoders; and 4) EchoFree~\cite{li2025echofree}, an ultra-lightweight hybrid framework that combines linear filtering with a Bark-scale neural post filter.
\subsubsection{Evaluation Metrics}
We evaluate performance using AECMOS~\cite{purin2022aecmos}, including EchoMOS for echo suppression and DegMOS for speech quality, together with ERLE for echo attenuation. PESQ~\cite{rix2001perceptual} and STOI~\cite{taal2011algorithm} are used to measure speech quality and intelligibility, respectively.

\tabvspace
\subsection{Bark- vs.\ Frequency-Domain Modeling}

Table~\ref{tab:compact_ablation} compares Bark- and frequency-domain models under different bottleneck channel size $C$ and hidden size $H$, with Fig.\,\ref{fig:compact_latent} showing Pareto efficiency and compression sensitivity. 
\cblue{The results show that frequency-domain models consistently outperform Bark-domain models under the same $(C,H)$ setting, but require much higher FLOPs. Nevertheless, the performance gap is considerably smaller in EchoMOS than in PESQ and ESTOI. This suggests that Bark-domain compression preserves the coarse spectral structures and energy patterns relevant to echo suppression, while reducing the spectral resolution available for modeling fine speech characteristics, which may contribute to degraded speech quality and intelligibility.
Considering footprint, Bark-domain models dominate the low- and middle-complexity regions of the Pareto frontier, while frequency-domain modeling achieves the highest score at much higher FLOPs. }
Fig.,\ref{fig:compact_latent}(b) shows that Bark-domain models are more sensitive to channel compression than hidden-dimension reduction. Reducing $C$ from 64 to 32 increases the normalized drop from 2.47\% to 5.50\%, whereas further reducing $H$ from 64 to 32 increases it only to 7.70\%.

\tabvspace
\subsection{Ablation Study of EAM}

\begin{figure}[t]
	\setlength{\abovecaptionskip}{-0.2cm}
	\centering
	\includegraphics[width=0.95\linewidth]{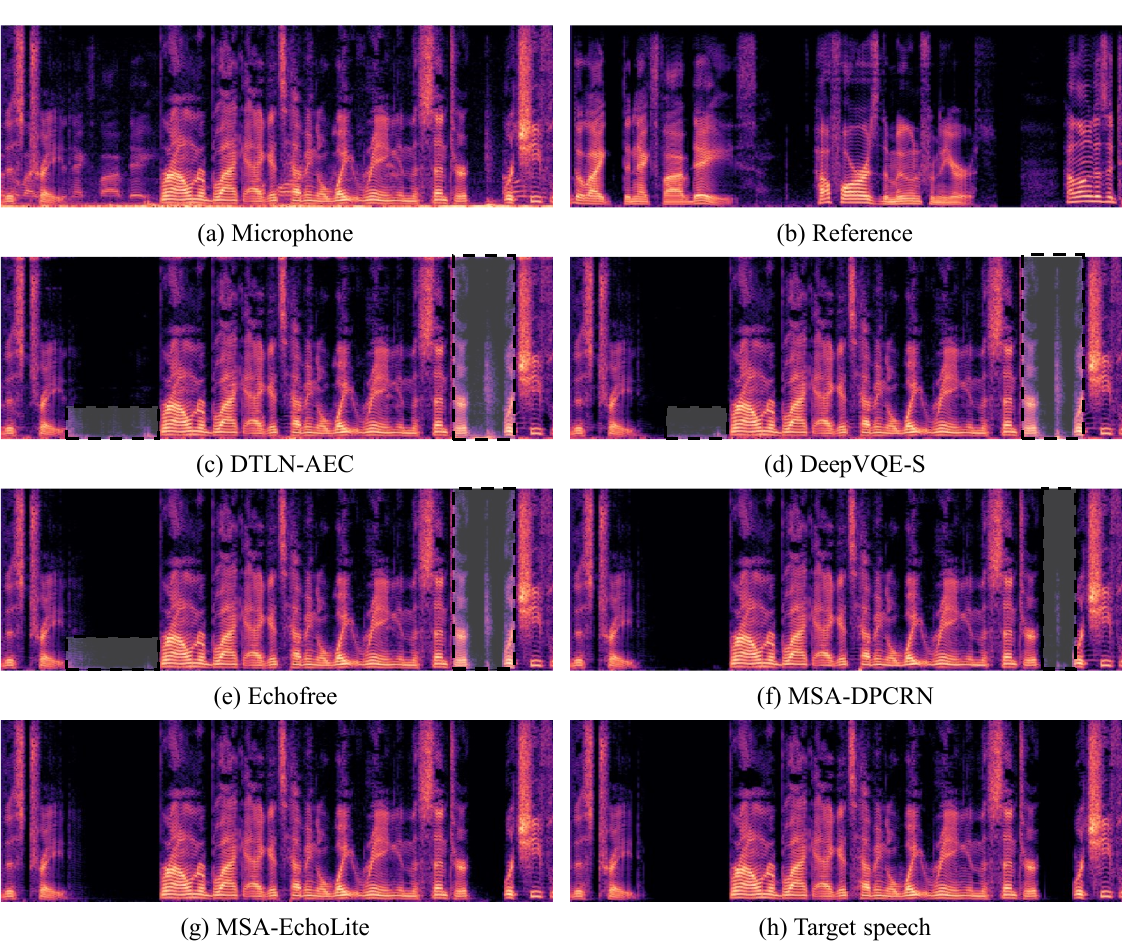}
\caption{Spectrogram comparison on a double-talk test sample.}
	\label{fig:spec}
	\vspace{-0.2cm}
\end{figure} 

We evaluate EAM under different compact-latent configurations. 
Table~\ref{tab:compact_ablation} and Table~\ref{tab:compact_ablation_2} show that +EAM consistently improves all configurations, confirming its effectiveness for compact-latent FT modeling. 
+EAM achieves the most pronounced improvement at C32H64, where the baseline suffers the largest degradation.
At this setting, EAM improves PESQ-WB, ESTOI, SDR, and DegMOS from 2.02, 79.67\%, 11.59\,dB, and 3.84 to 2.10, 80.68\%, 11.93\,dB, and 3.94, respectively, with a moderate increase in parameters from 0.15 to 0.20\,M and FLOPs from 80.83 to 101.90 M/s.
\cblue{Similar trends are observed across all Bark-band settings, with +EAM consistently improving the Bark-domain baselines in all evaluated metrics.}
These results indicate that EAM partially alleviates the degradation caused by compact bottleneck representations with limited additional complexity.

\tabvspace
\subsection{Comparison with Baseline Methods}

Table~\ref{table_ablation_study_dns} compares MSA-EchoLite with representative AEC baselines on the real blind test sets of the AEC Challenge. 
MSA-EchoLite achieves the best EchoMOS scores across all scenarios, improving the DT and FE EchoMOS to 4.28 and 4.26, respectively. 
Specifically, MSA-EchoLite achieves higher EchoMOS and comparable ERLE than MSA-DPCRN with much lower FLOPs (101.90\,M/s vs.\ 297.38\,M/s).
Compared with EchoFree, MSA-EchoLite achieves stronger echo suppression, especially in noisy and FE scenarios, at a moderate computational cost.
The results also reveal a trade-off between echo suppression and speech preservation, reflected by EchoMOS and DegMOS. 
Although PBFDAF and EchoFree achieve competitive DegMOS in some scenarios, PBFDAF's low EchoMOS indicates severe residual echo.
In contrast, MSA-EchoLite provides a better overall balance between echo suppression and speech preservation.
Fig.\,\ref{fig:spec} further visualizes that MSA-EchoLite better suppresses residual echo while preserving speech structures.

\tabvspace
\section{Conclusion}

\cblue{In this work, we proposed MSA-EchoLite, a lightweight hybrid AEC framework for resource-constrained scenarios.
Following the hybrid AEC paradigm, MSA-EchoLite uses PBFDAF to obtain an echo-aligned reference and performs Bark-domain neural suppression.
The results show that Bark-domain processing offers a better performance-complexity trade-off than frequency-domain processing, but is more sensitive to feature compression.
The proposed EAM-FT block mitigates the degradation caused by this compression by introducing echo-aware interaction cues.
Overall, MSA-EchoLite achieves stronger echo suppression than state-of-the-art low-complexity AEC baselines with a small computational footprint.
In the future, we will explore more efficient compact-domain bottleneck frequency-time modeling methods with lower sensitivity to the channel dimension.}

 \newpage

\bibliographystyle{IEEEtran}
\bibliography{main}

\end{document}